\documentstyle[twocolumn,prl,aps]{revtex}
\newcommand{\be}{\begin{equation}} \newcommand{\ee}{\end{equation}} 
\newcommand{\bea}{\begin{eqnarray}}\newcommand{\eea}{\end{eqnarray}}
\newcommand{\bm}[1]{\mbox{\boldmath$#1$}}
\newcommand{\grad}{\bm \nabla}
\begin{document}
\draft
\preprint{OCHA-SP-01-05, cond-mat/0109073}
\title{ Explosion-implosion duality in the Bose-Einstein condensation}
\author{Pijush K. Ghosh$^{*}$}
\address{ 
Department of Physics,
Ochanomizu University,\\
2-1-1 Ohtsuka, Bunkyo-ku,
Tokyo 112-8610, Japan.\\}

\maketitle
\begin{abstract} 
We show an explosion-implosion duality in the one and two dimensional
Bose-Einstein condensation with or without a particular time-dependent
harmonic trap. The result is independent of the strength and the attractive
or the repulsive nature of the self-interaction of the condensate. This implies
that the implosion in a particular atomic species has a dual description in
terms of the explosion in the same or another atomic species and the vice
versa. The result is applicable without any modifications to non-relativistic 
theories that are invariant under the $SL(2,R)$ reparameterization of the
time-coordinate.
\end{abstract}
\narrowtext

Duality plays an important role in science. Models with a dual description
are also abundant in different branches of physics. The duality in the two
dimensional Ising model\cite{ising}, the electromagnetic duality in
supersymmetric gauge theories\cite{sen}, the explosion-implosion duality in
astrophysics\cite{astro,astro1} are a few celebrated examples among them. The
purpose of this letter is to show an explosion-implosion duality in the
Bose-Einstein condensation (BEC) within the framework of the Gross-Pitaevskii
equation(GPE).

The action describing the two dimensional BEC without the harmonic trap is
invariant under a $SL(2,R)$ reparameterization of the time coordinate, leading
to a dynamical $O(2,1)$ symmetry of the model \cite{me}. The same is true
for the one dimensional BEC with repulsive interaction\cite{me}. One special
case of the $SL(2,R)$ reparameterization of the time coordinate corresponds
to the inversion, $ \tau \rightarrow t = - \frac{1}{\tau}$. The field and the
space-coordinates also transform simultaneously\cite{me} to keep the action
invariant. This particular symmetry, which we refer to as the duality-symmetry,
is the central to the discussions of this letter. We show that the
duality-symmetry leads to explosion-implosion duality in the GPE without
the trap.

The presence of the harmonic trap is essential for many experimental
observations in BEC. Thus, it is natural to study if the duality-symmetry
discussed above can be preserved in the BEC even in presence of the harmonic
trap. Unfortunately, the introduction of a time-dependent harmonic trap breaks
both the scale invariance and the invariance under a translation in time. This
leads to breaking of the dynamical $O(2,1)$ symmetry. We, nevertheless, show
that the duality-symmetry is present for a very particular choice of the
time-dependent harmonic trap. The time-dependent frequency of this particular
trap, which varies inverse-squarely with time, is independent of the space
dimensionality and the same for both the one( repulsive case only ) and the two
dimensional BEC.

We further show that the invariance under the duality-symmetry leads to
explosion-implosion duality in the one( repulsive case ) and two dimensional
GPE with the particular time-dependent trap. In particular, following the
method of Ref. \cite{me}, we first obtain
the exact time-dependence of the width of the wave-packet in the physical
problem described in terms of $\tau$. We also obtain the same in the dual
problem described in terms of variables obtained from the physical model by
the duality-transformation. The width in the physical problem is zero
initially. The width spreads gradually with the increase of the time and
finally diverges at the asymptotic infinity. On the other hand, in its
dual description, the
width diverges at $t=-\infty$. As $t$ increases, the width first takes a finite,
but, large value and then decreases with the passage of time. The width finally
vanishes at $t=0$.  Thus, the explosion in the physical problem has a dual
description in terms of implosion in the dual problem, leading to an
explosion-implosion duality in the BEC. We remark here that, unlike in the
case of wave collpase at a finite time\cite{rev}, the explosion/implosion
occurs only at the final value of the allowed ranges of the time coordinate
$\tau$/$t$. This difference is very crucial in understanding our results.

The consequences of such an explosion-implosion duality is the following. 
The self-interaction in the GPE can be either attractive or repulsive
depending on whether the s-wave scattering length of the atomic species
is negative or positive, respectively\cite{bec}. Further, using the Feschbach
resonance method\cite{fbr}, the magnitude and the sign of the self-interaction
between atoms of a certain species can be tuned to any desired value : large or
small, repulsive or attractive. If the self-interaction is
repulsive and sufficiently strong so as to overcome the effect of the confining
potential, the condensate is expected to explode. On the other
hand, we expect the condensate to implode in the strongly attractive regime
of the self-interaction, if the zero-point energy due to the kinetic energy 
term can no longer balance the attraction. However, contrary to this known
behavior of the condensate, we show that the exact
explosion-implosion duality is independent of the strength and the attractive
or the repulsive nature of the self-interaction. Thus, it is possible to
have both explosion and implosion for any fixed value of the scattering length
of an atomic species, provided the condensate is prepared satisfying the
required initial conditions described below. The reason behind such an
interesting, novel and apparently counter-intuitive prediction is that there is
no scale in the problem to prevent an initially growing/collapsing condensate
from further growth/collapse. The explosion in an atomic species can also be
described in terms of an implosion at a different value of the scattering
length of the same or another atomic species and the vice-versa.

Consider the following non-relativistic Lagrangian in arbitrary $d+1$
dimensions,
\be
{\cal{L}} = i \psi^* \partial_{\tau} \psi
- \frac{1}{2 m} {\mid \grad \psi \mid}^2
- g V({\mid \psi \mid}, {\bf r}).
\label{eq0}
\ee
\noindent The coupling constant $g$ has the inverse-mass dimension in the
natural units with $c=\bar{h}=1$. The potential $V$ is real and does not
depend on any dimensional coupling constant. This implies a scale-invariance
in the theory. We demand the invariance of the action
${\cal{A}} =\int d \tau d^d{\bf r} {\cal{L}}$ under the following
time-dependent transformations\cite{kl,trans,tala},
\bea
&& {\bf r} \rightarrow {\bf r_h} = {\dot{\tau}}(t)^{-\frac{1}{2}} {\bf r}, \ \
\tau \rightarrow t = t(\tau), \ \ \dot{\tau}(t) =
\frac{d \tau(t)}{d t}, \nonumber \\
&& \psi(\tau, {\bf r}) \rightarrow \psi_h(t, {\bf r}_h) =
\dot{\tau}^{\frac{d}{4}} exp \left ( - i m \frac{\ddot{\tau}}{4 \dot{\tau}}
r_h^2 \right ) \psi(\tau, {\bf r}),
\label{eq1}
\eea
\noindent with the scale-factor $\tau$ given by,
\be
\tau(t) = \frac{ \alpha t + \beta}{\gamma t + \delta}, \ \
\alpha \delta - \beta \gamma =1.
\label{eq2}
\ee
\noindent Particular choices of $\tau(t)=  t + \beta, \alpha^2 t$,
and $ \frac{t}{1 + \gamma t}$, correspond to time translation, dilation
and special conformal transformation (SCT). The Noether charges corresponding
to these symmetry transformations close under an $O(2,1)$ algebra\cite{me}.

The action ${\cal{A}}$ is invariant under the transformation (\ref{eq1}) and
(\ref{eq2}) for $g=0$\cite{trans}. For $g \neq 0$, the invariance of
${\cal{A}}$ under (\ref{eq1}) and (\ref{eq2}) solely depends on the form of $V$.
We choose,
\be
V ({\mid \psi \mid} ) = {\mid \psi \mid}^{\frac{4}{d} + 2}.
\label{eq3}
\ee
\noindent This choice of $V$ gives the critical non-linear Schr$\ddot{o}$dinger
equation(NLSE) in $d+1$
dimensions\cite{rev}. The Lagrangian (\ref{eq0}) with the above choice of
the interaction has an $O(2,1)$ symmetry for arbitrary $d$. Note
that $d=1$ gives a sextic interaction that is relevant for one dimensional
BEC with the repulsive interaction\cite{kolo}. The BEC in $2+1$ dimensions
is described by a quartic nonlinear interaction\cite{bec}. Note that for
$d=2$, we indeed have $V={\mid \psi \mid}^4$. This produces a cubic NLSE which
is also directly relevant for two dimensional optics\cite{rev,tala}. As an
aside, we remark here that for $d=3$, the equivalent formulation of (\ref{eq0})
and (\ref{eq3}) in terms of hydrodynamic variables \cite{fr} is directly
relevant for the supernova explosion or implosion in laser induced
plasma\cite{astro}.

The Lagrangian (\ref{eq0}) with the potential given by (\ref{eq3}) is invariant
under the transformations (\ref{eq1}) and (\ref{eq2}). Consider a very
particular case of Eq. (\ref{eq2}),
\be
\alpha=\delta=0, \ \ \gamma = - \frac{1}{\beta}, \ \ 
\tau = - \frac{\beta^2}{t}.
\label{eq4}
\ee
\noindent This implies the following symmetry-transformations for the
field $\psi$ and the coordinate ${\bf r}$,
\bea 
&& {\bf r} \rightarrow {\bf r_h} = \frac{t}{\beta} {\bf r} =
 - \frac{\beta}{\tau} {\bf r}, \nonumber \\
&& \psi(\tau, {\bf r}) \rightarrow \psi_h(t, {\bf r}_h) =
\left ( \frac{\beta}{t} \right )^{\frac{d}{2}}
exp \left ( i \frac{m t}{2 \beta^2} r^2 \right ) \psi(\tau, {\bf r}),
\label{eq4.0}
\eea
\noindent which is known as the lens transformation in the context of
the critical NLSE\cite{rev,tala}. The parameter $\beta$ is real and arbitrary.
Consequently, the theory at a time
$\tau > 0 $ is mapped to a theory at a time $ t < 0$ and the vice versa. In
particular, the physical problem at $\tau = 0$ is mapped to the dual problem at
$t = - \infty$. Similarly, the physical problem at $\tau = \infty$ is mapped
to the dual problem at $t = 0$. The critical value separating this two regime
is obviously given by $\tau=0$ or $t=0$. Thus, the physical problem is bounded
from below in time, while the dual problem is bounded from above in time. The
above mapping also involves a time-dependent scaling for the space-coordinate.
We choose the scale-factor $\frac{t}{\beta} = - \frac{\beta}{\tau}$ to be
positive-definite and follow the convention for the time-coordinates and the
parameter $\beta$ as given below,
\be
0 \leq \tau \leq \infty, \ \ - \infty \leq t \leq 0, \ \ 
\beta < 0.
\label{eq4.1}
\ee
\noindent Note that $\beta$ has been chosen to take
negative values only. If one prefers to choose a positive value for $\beta$,
the ranges of the time-coordinates in (\ref{eq4.1}) should also be interchanged
simultaneously.
We remark that the positivity of the scale-factor is essential
for the consistency of our analysis. For example, the densities
$\rho=\psi^* \psi$ and $\rho_h=\psi^*_h \psi_h$ are related to each other
through the following equation,
\be
\rho_h(t, {\bf r}_h ) = \left (-\frac{\tau}{\beta} \right )^d 
\rho(\tau, {\bf r}) = \left ( \frac{\beta}{t} \right )^d \rho(\tau, {\bf r}).
\label{eq4.11}
\ee
\noindent This is consistent for odd $d$ only if the scale-factor is
positive-definite. Further, the width of the wave-packet ( see below ) in
arbitrary $d$ becomes negative, unless we demand the positivity of the
scale-factor.

Let us now introduce a moment $I$ and its dual ${\cal{I}}$ as,
\be
I (\tau) = \frac{m}{2} \int d^d {\bf r} \ r^2 \rho, \ \
{\cal{I}}(t) = \frac{m}{2} \int d^d {\bf r}_h \ r_h^2 \rho_h.
\label{eq4.12}
\ee
\noindent The moment $I$ and its dual ${\cal{I}}$ can be interpreted as
the expectation value of the square of the radius of the condensate.
Using the transformation (\ref{eq4.0}), it is easy to see that the
moment $I$ and its dual ${\cal{I}}$ are related to each other by the
equation,
\be
{\cal{I}} (t) = \left ( \frac{t}{\beta} \right )^2 I(-\frac{\beta^2}{t}).
\label{eq4.12.5}
\ee
\noindent Following Ref. \cite{me} and Eq. (\ref{eq4.12.5}), $I(\tau)$ and
${\cal{I}}(t)$ are universally given by,
\be
I = ( a + b \tau )^2 + \frac{Q}{a^2} \tau^2, \ \
{\cal{I}} = \left ( b \beta - \frac{a}{\beta} t \right )^2 +
\frac{Q \beta^2}{a^2},
\label{eq4.13}
\ee
\noindent where $a$ and $b$ are arbitrary constants. The constant of motion
$Q= I H - (\frac{1}{2} \frac{d I}{d \tau} )^2$ is related to the
Casimir operator of the underlying $O(2,1)$ symmetry\cite{me}. Note
that all the information on the system under
consideration ( like the strength and the attractive or the repulsive nature
of the interaction, the space-time dimension on which the problem is
considered, the nonlinearity etc. ) is contained in the expression of
$I$ and ${\cal{I}}$ in the constant of motion $Q$ only, through the
Hamiltonian $H$. Thus, the dynamics can be described in terms of the same
set of initial conditions for any value of $g$: positive or negative, large
or small. It is possible to have both explosion and implosion for a particular
value of $g$, if the condensate is prepared following the respective initial
conditions, which can be extracted from the exact expressions of $I$ and
${\cal{I}}$. One might naively think this as unphysical. However, note that
both the kinteic energy term and the interaction term scales in the same
way. So, there is no scale in the problem to prevent an initially
growing/collapsing condensate from further growth/collapse.

The criteria for the collapse of the condensate at a finite and real time
$\tau^*$ is $Q \leq 0$. In particular, the moment $I$ vanishes at a finite
time $\tau^*$,
\be
\tau^* = \frac{a^2}{(a^2 b^2 + Q)} \left [ - ab \pm \sqrt{-Q} \right ],
\label{eq4.2.1}
\ee
\noindent which is real if $ Q \leq 0$. Note that we have the freedom of
making $\tau^*$ either positive or negative by choosing appropriate values
for the integration constants $a$ and $b$. The moment $I$ is semi-positive
definite by definition. Thus, the exact expression for $Q=I H -
(\frac{1}{2} \frac{d I}{d \tau})^2$ implies that the condensate collapses for
any initial condition if $ H \leq 0$. On the other hand, if $H > 0$, the
condition for the collapse is given by,
$\frac{d I}{d \tau} \Bigm|_{\tau=0} \leq - 2 \sqrt{I\mid_{\tau=0} {\mid
H \mid}}$. Thus, the explosion-implosion duality is forbidden for $Q \leq 0$,
since there can not have any explosion either in the physical
or in the dual theory.

We now consider the case $Q > 0$ for which the explosion-implosion duality
can be realized. A positive $Q$ necessarily implies a positive $H$,
$ H \geq \frac{1}{I} (\frac{1}{2} \frac{d I}{d \tau} )^2$.
This lower bound on $H$ can be equivalently written as a constraint on
the initial profile of the condensate,
\be
H \geq  \left [ \frac{1}{4 I} \left ( \frac{d I}{d \tau} \right )^2 \right ]
\Bigm|_{\tau=0}, \ \
{\cal{H}} \geq \left [ \frac{1}{4 {\cal{I}}} \left ( 
\frac{d {\cal{I}}}{d t} \right )^2 \right ]
\Bigm|_{t=-\infty}, \ \
\label{eqcri}
\ee
\noindent where ${\cal{H}}$ is the dual Hamiltonian. Using the exact
time-dependence of $I$, ${\cal{I}}$ and an alternative expression
for $H({\cal{H}})$, $H = \frac{1}{2} \frac{d^2 I}{d \tau^2}
( {\cal{H}} = \frac{1}{2} \frac{d^2 {\cal{I}}}{d t^2})$, it is easy to verify
that such a criteria is indeed satisfied. Finally, whether a condensate
satisfying the criteria (\ref{eqcri}) will explode or implode depends
on whether the expectation value of the square of the radius of the
condensate is finite or infinite, respectively, at the initial time.

The condensate in the physical problem is of finite extent at $\tau=0$ with
$I(0)=a^2$. The condensate swells gradually as $\tau$ increases and
explodes at $\tau=\infty$ leading to a divergence in $I$. In the dual
description, the condensate is of infinite extent at $t=-\infty$ with a
divergent ${\cal{I}}$. As $t$ increases, the condensate is extended
over a very large, but, finite area. It then shrinks with the advancement
of time, finally reaching to a constant value for the moment ${\cal{I}}$
at $t=0$, ${\cal{I}}(0) = \beta^2 \left ( b^2 + \frac{Q}{a^2} \right )$.
Thus, the evolution of the condensate in the physical problem is different
from that of its dual description. In particular, the explosion in the physical
problem has a dual description in terms of implosion. Note that we have the
freedom of choosing $ a^2 > \sqrt{Q} {\mid \beta \mid}, b^2 = (a \beta)^{-2}
( a^4 - Q \beta^2)$, such that, ${\cal{I}} (0) = I ( 0 ) = a^2 $. This
allows the condensates to have the same radius, both at the initial time in
the physical problem and at the final time in the dual description.

A comment is in order at this point. The lens transformation (\ref{eq4.0}) has
the property of converting a collapsing solution to a stationary solution
known as the Townes soliton\cite{rev,tala}. The Hamiltonian vanishes, when
evaluated for the Townes soliton\cite{rev}, implying that the constant of
motion $Q$ is necessarily negative. So, there is no contradiction between
the known results on the Townes solition and the results of the present
letter on the explosion-implosion duality which is strictly valid for
$Q \geq 0$.

We now show that it is possible to choose a particular time-dependent
harmonic trap maintaining the duality-symmetry. Consider the following
Lagrangian,
\be
{\cal{L}}_h = {\cal{L}} - \frac{1}{2} m \omega(\tau) r^2 {\mid \psi \mid}^2,
\ \ \omega(\tau) = \omega_0^2 \left ( \gamma \tau - \alpha \right )^{-2},
\label{eq5}
\ee
\noindent where we have made a very specific choice for the frequency of the
time-dependent trap and $\omega_0$ is an arbitrary constant. The action
${\cal{A}}_h = \int d\tau d^d {\bf r} {\cal{L}}_h$ is
invariant under the transformation (\ref{eq1}) with the $\tau(t)$ given by,
\be
\tau(t) = \frac{ \alpha t + \beta}{\gamma t - \alpha}, \ \
\alpha^2 + \beta \gamma = - 1.
\label{eq6}
\ee
\noindent Note that the time-translation, the dilatation and the special
conformal transformation can not be obtained as special cases from the
above relation. However, the transformation responsible for the duality-symmetry
is contained in (\ref{eq6}). This is obtained by putting $\alpha=0$ and,
hence, $\gamma = - \frac{1}{\beta}$,
\be
\omega(\tau) = \left ( \frac{\omega_0 \beta}{\tau} \right )^2. 
\label{eq6.1}
\ee
\noindent For this choice of the time-dependent frequency, the action
${\cal{A}}_h$ is invariant under the transformations given by the Eqs. 
(\ref{eq4}) and (\ref{eq4.0}). Thus, although the action ${\cal{A}}_h$ is
not invariant under the individual time-translation or the SCT,
it is indeed invariant under the duality-symmetry.

Using the transformations (\ref{eq1}) and (\ref{eq2}), the action ${\cal{A}}$
can be mapped\cite{me} to the action ${\cal{A}}_h$ with the time-dependent
frequency given by (\ref{eq6.1}), if we choose,
\be
\tau(t) = \frac{1}{c_2 ( 2 \eta - 1 )} \left ( c_1 + 
c_2 \ t^{1 - 2 \eta} \right ) ^{-1},
\label{eq6.2}
\ee
\noindent where $c_1$ and $c_2$ are arbitrary constants. The parameter $\eta$
is determined in terms of $\beta$ and $\omega_0$,
\be
\eta = \frac{1}{2} \left [ 1 + ( 1 - 
4 \omega_0^2 \beta^2 )^{\frac{1}{2}} \right ].
\label{eq6.3}
\ee
\noindent For ${\mid \omega_0 \mid} \geq \frac{1}{2 {\mid \beta \mid}}$,
$\eta$ becomes a complex number. Moreover, at ${\mid \omega_0 \mid} =
\frac{1}{2 {\mid \beta \mid}}$, $\tau(t)$ becomes divergent. Thus, we restrict
$\omega_0$ to take values within the range,
$ 0 \leq {\mid \omega_0 \mid} < \frac{1}{2 {\mid \beta \mid}}$.
Consequently, $\eta$ is restricted to as, $ \frac{1}{2} < \eta \leq 1 $. Note
that ${\cal{A}}_h$ is invariant under the transformations (\ref{eq4}) and
(\ref{eq4.0}) for arbitrary $\omega_0$. The restriction on $\omega_0$ is
valid only if we want to relate ${\cal{A}}$ to ${\cal{A}}_h$ through
a time-dependent coordinate transformations. The BEC without the trap
can be obtained in the limit $\omega_0 \rightarrow 0$ or $\eta \rightarrow 1$.
The transformation (\ref{eq6.2}) reduces to a symmetry transformation
for $\eta=1$ and keeps ${\cal{A}}$ invariant.
 
The moment $I$ for the BEC with the harmonic trap can be interpreted as
the square of the width of the wave-packet\cite{spain}.
Following \cite{me}, the moment $I$ and its dual ${\cal{I}}$ for the
trapped BEC are determined as,
\bea
&& I^h = \tau^{ 2 \eta} +
\frac{Q}{(1 - 2 \eta)^2} \tau^{ 2 ( 1 - \eta )},\nonumber \\
&& {\cal{I}}^h = {\mid \beta \mid}^{ 2 ( 2 \eta - 1 )}
\left ( - t \right )^{ 2 ( 1 - \eta )} + \frac{ Q {\mid \beta \mid}^{2 ( 1 -
2 \eta ) }}{(1 - 2 \eta)^2} ( - t )^{2 \eta}.
\label{eq6.4}
\eea
\noindent We have chosen 
$ a c_2 = \pm \sqrt{Q} \left ( 1 - 2 \eta \right )^{-1}$ and
$ \frac{b}{c_1} = \pm \sqrt{Q} $
in deriving Eq. (\ref{eq6.4}).
Recall that the coordinate $t$ is allowed to take values in
the negative-axis only and $Q \geq 0$.

The width $I^h$ vanishes at $\tau=0$ in the physical problem. This is because
the confinement is too strong to allow any non-vanishing width in this
limit. The strength of the confinement decreases with the increase of
$\tau$ and the width starts spreading over the accessible area. This gradually
leads to a divergence in the width at the asymptotic infinity, where the
confinement is so weak that it can not hold the condensate any longer.
This signifies an explosion. In the dual description of this problem, the
width ${\cal{I}}^h$ diverges at $t=-\infty$. As $t$ increases, the width
first becomes large, but, finite and then decreases gradually. This leads to a
collapse of the condensate with vanishing width at $t=0$. This signifies an
implosion. Thus, the explosion in the physical problem has the dual description
in terms of implosion.

In conclusion, we have shown the explosion-implosion duality in the 
one(repulsive case) and two dimensional BEC with or without the particular
time-dependent harmonic trap. This implies that the implosion in a particular
atomic species has the dual description in terms of the explosion in the same
or another atomic species and the vice versa. With the recent advancement
of the technology related to the BEC\cite{fbr,ld}, such a duality may be
realized in
the laboratory in near future. Finally, we would like to mention that {\it the
method and the results of this letter are applicable without any modifications
to non-relativistic theories that are invariant under the $SL(2,R)$
reparameterization of the time coordinate}. Physically interesting models with
such an invariance are abundant\cite{exa0,exa1,exa2}. It is surprising to note
how the underlying dynamical $O(2,1)$ symmetry gives a universal result for
models ranging from astrophysics to models as diverse as BEC.

\acknowledgements{I would like to thank T. Deguchi and T. K. Ghosh for their
interest in this work. This work is supported by a fellowship
(P99231) of the Japan Society for the Promotion of Science.}

\end{document}